\documentclass[prd,twocolumn,superscriptaddress,showpacs,aps]{revtex4}
\usepackage{graphicx} 
\usepackage{bm} 

\newcommand{\beq}{\begin{equation}}
\newcommand{\eeq}{\end{equation}}
\newcommand{\bea}{\begin{eqnarray}}
\newcommand{\eea}{\end{eqnarray}}
\newcommand{\ba}{\begin{array}}
\newcommand{\ea}{\end{array}}
\newcommand{\lsim}   {\mathrel{\mathop{\kern 0pt \rlap
  {\raise.2ex\hbox{$<$}}}
  \lower.9ex\hbox{\kern-.190em $\sim$}}}
\newcommand{\gsim}   {\mathrel{\mathop{\kern 0pt \rlap
  {\raise.2ex\hbox{$>$}}}
\lower.9ex\hbox{\kern-.190em $\sim$}}}

\begin{document}

\title{Second dip as a signature of ultrahigh energy proton
  interactions with cosmic microwave background radiation}

\author{V.~Berezinsky}
\affiliation{INFN, Laboratori Nazionali del Gran Sasso, I-67010
  Assergi (AQ), Italy}
\affiliation{Institute for Nuclear Research of the RAS, Moscow, Russia}
\author{A.~Gazizov}
\affiliation{B.~I.~Stepanov Institute of Physics of the National
  Academy of Sciences of Belarus, BY-220072 Minsk, Belarus}
\author{M.~Kachelrie\ss}
\affiliation{Institut for fysikk, NTNU, N-7491 Trondheim, Norway}

\begin{abstract}
We discuss as a new signature for the interaction of extragalactic
ultrahigh energy protons with cosmic microwave background radiation 
a spectral feature located at $E=6.3\times 10^{19}$~eV
in the form of a narrow and shallow dip. It is produced by the
interference of $e^+e^-$-pair and pion production. 
 We show that this dip and in particular its position are almost
model-independent. Its observation by future ultrahigh
energy cosmic ray detectors may give the conclusive confirmation that 
an observed steepening of the spectrum is caused by the
Greisen-Zatsepin-Kuzmin effect.

\end{abstract}
      
\maketitle

{\em Introduction.---}%
The nature and the sources of ultrahigh energy cosmic rays (UHECRs)
are not yet established despite more than 40~years of research.
Natural candidates as UHECR primaries are extragalactic protons from
astrophysical sources. In this case, interactions of UHE protons with 
the cosmic microwave background (CMB) leave their imprint on the UHECR 
energy spectrum in the form of the Greisen-Zatsepin-Kuzmin (GZK) 
cutoff~\cite{GZK} and a dip~\cite{HS85,BG88,Stanev00,BGG-dip}. 

The GZK cutoff is a steepening of the proton spectrum at the energy 
$E_{\rm GZK} \approx (4 - 5)\times 10^{19}$~eV, caused by photo-pion
production on CMB. This is a very spectacular effect, but the
shape of this steepening is strongly model-dependent~\cite{BBO,BGG}. 
Thus the GZK suppression is difficult to distinguish from, e.g., a
cut-off due to the maximal acceleration energy in a source.
The {\em dip\/} is a spectral feature produced by $p+\gamma_{\rm CMB}
\to p+e^++e^-$ interactions. It is a faint feature,    
practically not noticeable when the spectrum is plotted in the most
natural way, $J_p(E)$ versus $E$. The dip becomes more pronounced in the
modification factor \cite{BG88} $\eta(E)=J_p(E)/J_p^{\rm unm}(E)$, 
where $J_p(E)$ is the spectrum calculated with all energy losses 
included, and $J_p^{\rm unm}(E)$ is the unmodified spectrum calculated with 
adiabatic energy losses only. The dip is clearly
seen in the energy-dependence of $\eta(E)$ and is reliably confirmed by
observational data~\cite{BGG,BGG-dip}. 

In this Letter, we demonstrate the existence of one more signature of
UHE protons interacting with the CMB, which we call the {\em second dip}. 
In many aspects it is  similar to the first dip. The first dip starts
at the energy $E_{\rm eq1}=2.3\times 10^{18}$~eV, where pair-production
energy losses become equal to those due to redshift. The second dip
starts at the energy $E_{\rm eq2}=6.0\times 10^{19}$~eV, where photo-pion
energy losses become equal to those due to $e^+e^-$-pair production.
Both features are not seen well when the UHECR spectrum is displayed in a
natural way. While the first dip becomes visible dividing  the
experimental spectrum by the unmodified spectrum  $J_p^{\rm unm}(E)
\propto E^{-\gamma_g}$, the second dip appears dividing by the smooth
{\em universal spectrum\/} (see below).

{\em Kinetic equation, Fokker-Planck (FP) equation, and 
 continuous energy loss (CEL)  approximation.---}%
We shall calculate the diffuse spectrum of UHE
extragalactic protons assuming a homogeneous source distribution 
and a power-law generation spectrum with spectral index $\gamma_g$. 
In the CEL approximation,
the density of UHE protons at the present time $t_0$ can
be calculated from the conservation of the number of protons as 
\beq
n_p(E,t_0)dE =\int_{t_{\min}}^{t_0}dt \: Q_{\rm gen}(E_g)dE_g \,,
\label{conserv}
\eeq
where $t$ is the cosmological time and $Q_{\rm gen}\propto E_g^{-\gamma_g}$
is the particle generation rate per unit comoving volume. We denote by
$E_g(E,t)$ the initial energy of a proton generated at the cosmological epoch
$t$, if its present ($t=t_0$) energy is $E$. The energy evolution
$E_g(E,t)$ can be  easily calculated from the known energy losses. 
The solution of Eq.~(\ref{conserv}) was explicitly obtained in 
Refs.~\cite{BG88,BGG} and for a homogeneous distribution
of sources it is called {\em universal spectrum\/}
because it does not depend on the mode of propagation, being the same
e.g. for rectilinear and diffusive propagation~\cite{AB}. 
The universal spectrum is obtained in CEL approximation. With higher 
precision the spectrum can be calculated using a kinetic equation,  
\bea
\lefteqn{
 \frac{\partial n_p}{\partial t} = -3 H(t) n_p + 
\frac{\partial}{\partial E}\left \{ \left [H(t)E+b_{\rm pair}(E,t)\right ]
n_p\right \} } 
\label{KE}
\\ - & &  \!\!\!  \!\!\!  
P(E,t)n_p+ \int_E^{E_{\max}} \!\!\!\!\!dE' P(E',E,t) n_p(E',t) +
Q_{\rm gen}(E,t) \,.
\nonumber 
\eea
Here, $n_p\equiv n_p(E,t)$, $H(t)$ is the Hubble parameter,  
$b_{\rm pair}(E,t)$ are the energy losses due to 
pair-production treated in the CEL approximation, $P(E,t)$ is the exit
probability from the energy interval $dE$ due to $p\gamma \to \pi X$ collisions, 
and $P(E',E,t)$ is the probability that a proton with energy $E'$
produces a proton with energy $E$ in a $p\gamma \to \pi X$ collision. 
Introducing $x=E/E'$ and expanding the regeneration term in Eq.~(\ref{KE})
in a Taylor series with respect to $(1-x)$, one obtains at order
$(1-x)$ the CEL equation with the universal spectrum as solution. 
Including also the $(1-x)^2$ terms, the FP equation
emerges as
\bea
\frac{\partial n_p}{\partial t} &=& -3 H(t) n_p +
\frac{\partial}{\partial E}\left \{\left [H(t)E+b_{\rm tot}(E,t)\right ]
n_p\right \} 
\nonumber\\ & + &
\frac{\partial^2}{\partial E^2}\left [ E^2D(E,t)n_p \right ]+
Q_{\rm gen}(E,t) \,. 
\label{FP}
\eea
Here, $b_{\rm tot}(E)$ is the sum of  pair-production and photo-pion energy
losses in the CEL approximation and $D(E,t)=(\delta E)^2/\delta t$ is the 
diffusion coefficient in momentum space.

\begin{figure}[ht]
\includegraphics[width=7.4cm,height=4.9cm]{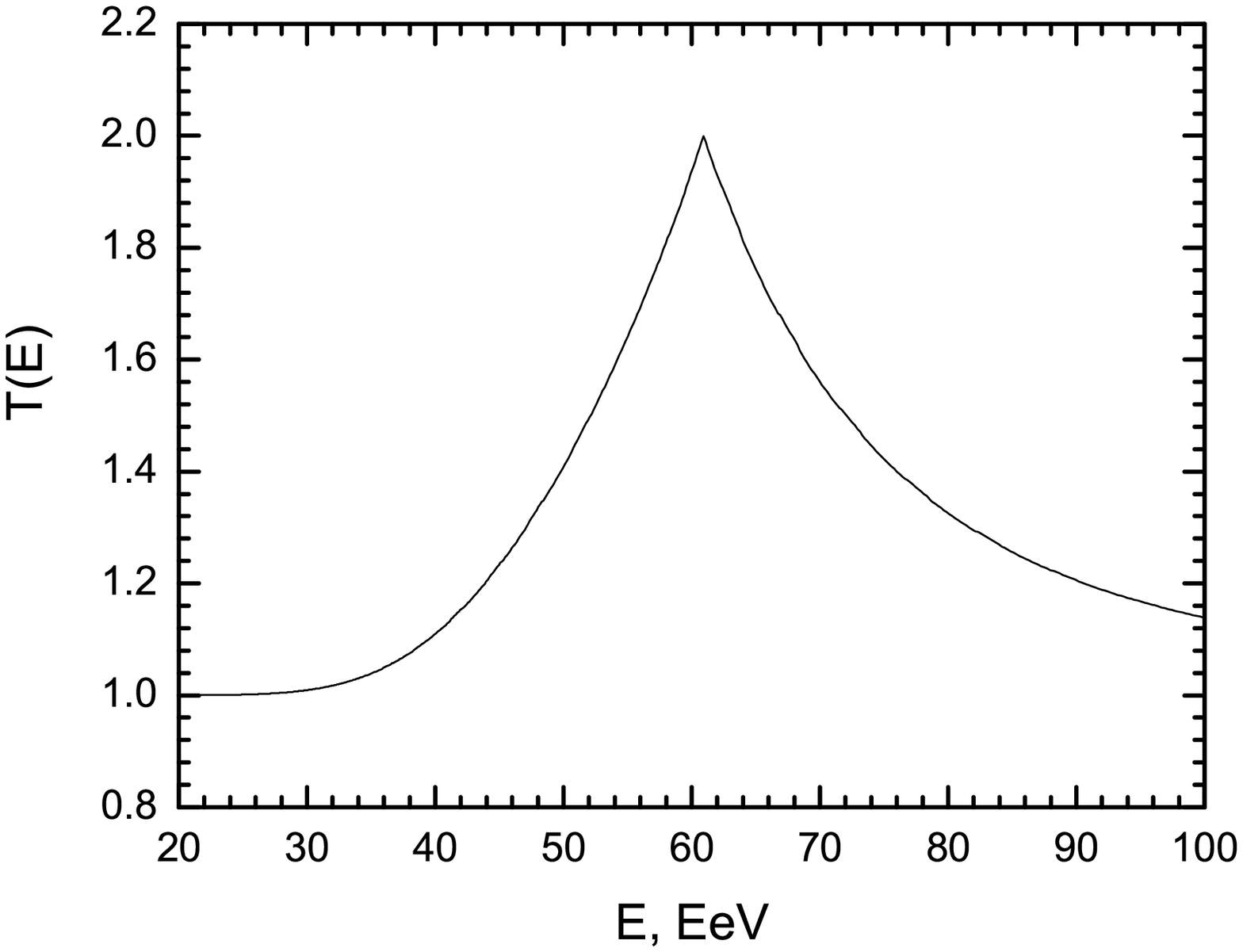}
\includegraphics[width=7.4cm]{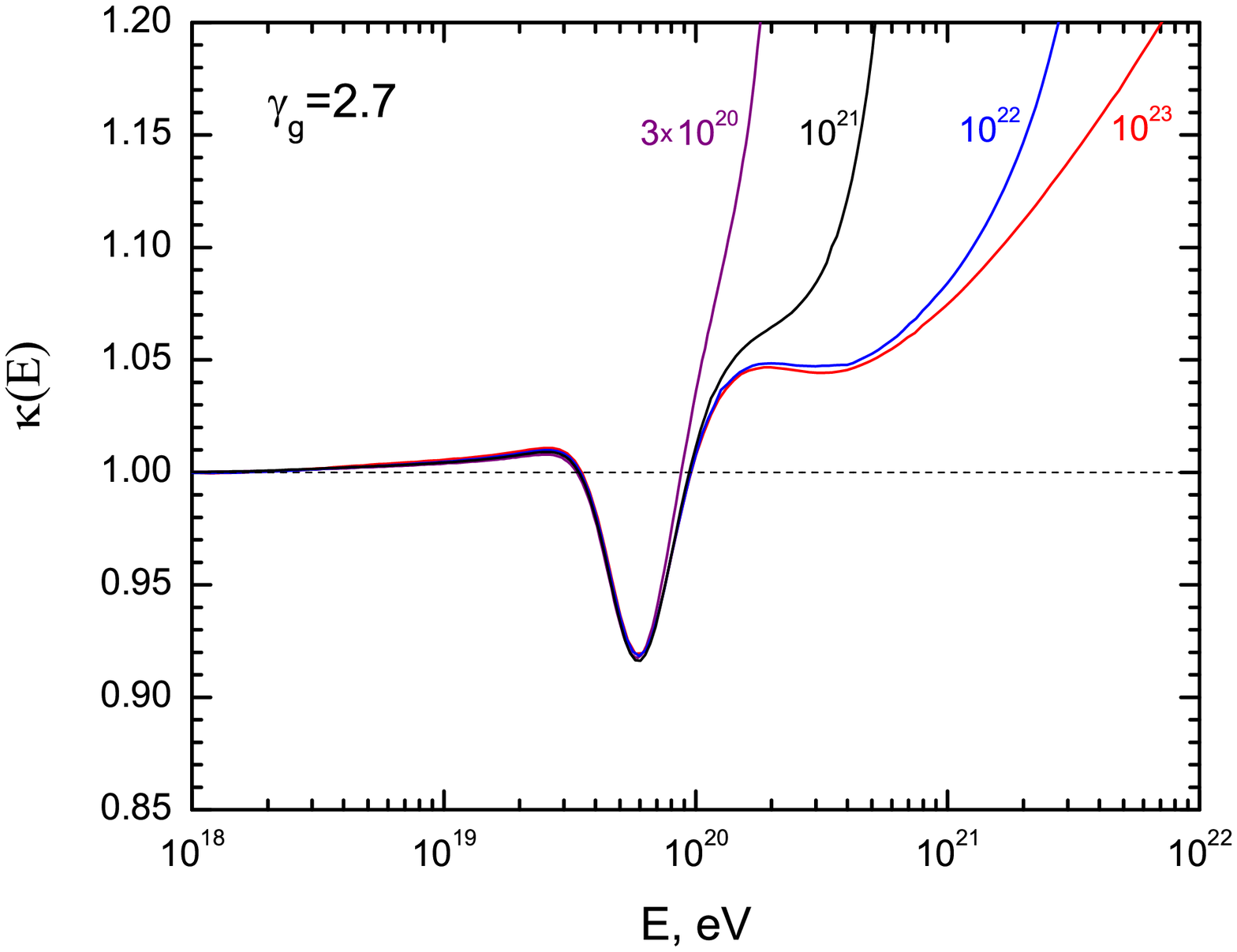}
\caption{Upper panel: the trigger function $T(E)$ as function
of energy. Lower panel: the distortion factor 
$\kappa (E)=J_{\rm kin}(E)/J_{\rm univ}(E)$ for $\gamma_g=2.7$ and 
different values of $E_{\rm max}$.}
\label{trigg}
\end{figure}

The kinetic equation~(\ref{KE}) allows us a transparent interpretation of
the spectral feature which appears at the energy 
$E_{\rm eq2}=6.0\times 10^{19}$~eV. At this energy one may 
limit the consideration to the present cosmological epoch $t \approx t_0$. 
As direct calculations show, the absorption term $-P(E)n_p(E)$ is then
compensated with high accuracy by the regeneration term with $P(E,E')$ in 
Eq.~(\ref{KE}). At $E \approx E_{\rm eq2}$, the small CEL term 
(pair production) breaks this compensation, increasing the absorption 
term, and the spectrum acquires a dip. It is quite narrow because  
photo-pion energy losses increase with energy almost exponentially and 
at $E > E_{\rm eq2}$ the pair-production energy losses become too
small. On the other hand, at $E< E_{\rm eq2}$ the photo-pion energy
losses are too small, the spectrum is fully determined by pair-production
energy losses, while the interference effect disappears. 

{\em Trigger mechanism.---}%
Prior to presenting exact numerical calculations we shall study
semi-quantitatively the {\em triggering mechanism}, responsible for the
second dip.  
One can rearrange the first three terms on the rhs of Eq.~(\ref{KE}) into 
$P_{\rm eff}(E,t)=P(E,t)+P_{\rm cont}(E,t)$ with 
\bea
\lefteqn{
P_{\rm cont}(E,t)  = 
2H(t) - \frac{\partial b(E,t)}{\partial E}}
\nonumber\\ & &- 
\left [\frac{b_{\rm pair}(E,t)}{E}+H(t) \right ]
\frac{\partial \ln n(E,t)}{\partial \ln E}  \,.
\label{P-cont}
\eea
It is the term $P_{\rm cont}(E,t)$ that breaks the above-mentioned
compensation between absorption and regeneration terms in
Eq.~(\ref{KE}), triggering thereby the modification of 
$n_{\rm kin}(E,t)$.

It is convenient to introduce  the auxiliary {\em trigger function\/}
$T(E)$  defined at $t=t_0$ as 
\beq
T(E)=\left\{ \begin{array}{ll}
P_{\rm eff}(E)/P_{\rm cont}(E)              ~&{\rm for}~~ E \leq E_c\\ 
P_{\rm eff}(E)/P(E)                         ~&{\rm for}~~ E \geq E_c ,
\end{array}
\right.
\eeq
where $E_c=6.1\times 10^{19}$~eV is determined from the condition 
$P(E)=P_{\rm cont}(E)$ and is approximately equal to 
$E_{\rm eq2}$.  The trigger function describes how 
$P_{\rm eff}(E)$ is changing from $P_{\rm cont}(E)$ at $E \ll E_c$, 
where $T(E)=1$, to $P(E)$ at $E \gg E_c$, where $T(E)=1$ as well. As
long as $T(E)\approx 1$, there is no interference between pair-production and
pion-production terms, and the ordinary solutions are valid. At $E=E_c$,
$T(E)$ reaches its maximum and $P_{\rm eff}(E_c)$, being noticeably larger
than $P(E)$, breaks the compensation between absorption and regeneration
terms in Eq.~(\ref{KE}), making absorption larger. As a result,
$n_{\rm kin}(E)$ decreases around $E_c$. The trigger function is plotted in
Fig.~\ref{trigg}. It reaches its maximum $T(E)=2$ at 
$E=E_c \approx 6.1\times 10^{19}$~eV. As explained above $n_{\rm kin}(E)$
must have a local minimum at this energy. The triggering mechanism
predicts that the position of the dip minimum $E_{\rm 2dip}$ does not 
depend on $\gamma_g$ and $E_{\rm max}$ and these predictions are
confirmed by our numerical calculations. The shape 
of $n_{\rm kin}(E)$ is expected to be similar to the shape of the
trigger function $T(E)$ and this expectation is also confirmed by
numerical calculations (see Fig.~\ref{trigg}). 

{\em Numerical solutions.---}%
We next discuss the second dip using numerical solutions 
of the kinetic equation (\ref{KE}). As mentioned above, the first dip is
distinctly seen in the energy dependence of the {\em  modification 
factor\/} $\eta (E)=J_p(E)/J_p^{\rm unm}(E)$. Similarly, the second dip 
is well seen, when the spectrum is described by a {\em distortion factor},
defined as $\kappa (E)=J_{\rm kin}(E)/J_{\rm univ}(E)$, where 
$J_{\rm univ}(E)$ is the universal spectrum from Eq.~(\ref{conserv}). We 
emphasize that the correct prediction for the measured spectrum
is given by the kinetic equation (\ref{KE}), while the universal
spectrum, used as a reference spectrum, is obtained in the CEL approximation
and as such does not include the interference between pair and pion
production. The calculated distortion factor is shown in the lower panel
of Fig.~\ref{trigg} for $\gamma_g=2.7$ and four values of $E_{\rm max}$.
The second dip is clearly seen. Its minimum is given by 
$E_{\rm 2dip}=6.3\times 10^{19}$~eV in good agreement with the
prediction of the triggering mechanism. The width of the dip also agrees
well with that of the trigger function $T(E)$. The independence of
the spectral shape of the dip from the numerical value of $E_{\rm
  max}$, seen in Fig.~\ref{trigg}, is another prediction of the 
triggering mechanism.
 
The distortion factor $\kappa$ does not return to unity after the second
dip, but continues to grow for $E\gg E_c$. 
This deviation from unity is explained by fluctuations in
photo-pion interactions. For $E \to E_{\rm max}$, the ratio 
$\kappa=J_{\rm kin}(E)/J_{\rm cont}(E) \to \infty$:  
At these energies  Eq.~(\ref{KE}) becomes stationary,
$$
- P(E)n_p(E)+ \int_E^{E_{\max}} \!\!\!\!
dE' P(E',E) n_p(E') + Q_{\rm gen}(E)=0 \,.
$$
When $E$ approaches $E_{\rm  max}$, the regeneration term disappears,
and one obtains  
$n_{\rm kin}(E_{\rm max})=Q_{\rm gen}(E_{\rm max})/P(E_{\rm max})$, remaining
finite at $E_{\rm max}$. Using the CEL approximation, the equation reads 
\beq
\partial/\partial E~ [b_{\rm tot}(E)n_p(E)]+Q_{\rm gen}(E)=0 \,.
\eeq
For $E \to E_{\rm max}$, $n_{\rm cont}(E) 
\propto (E_{\rm max}-E)/E_{\rm max} \to 0$ and hence 
$\kappa(E_{\rm max})=n_{\rm kin}(E_{\rm max})/n_{\rm cont}(E_{\rm max}) \to
\infty$. The explosive behavior of the distortion factor for 
$E\to E_{\rm max}$ reflects the different limiting density of $n_p(E)$
in the kinetic and CEL equations. This result has
a clear physical meaning. 
A particle has a finite probability to travel a finite distance
without losing energy. While the kinetic equation describes 
correctly this effect, in the CEL approximation particles lose energy 
for any distance traversed, however small it may be. This influences the ratio 
$n_{\rm kin}(E)/n_{\rm cont}(E)$ at all energies close enough to
$E_{\rm max}$ as can be seen in Fig.~\ref{trigg}.

We have obtained one more proof for the energy-losses interference as 
the origin of the second dip. For this we have calculated the  
distortion factor in a toy model in which 
pair production  and adiabatic energy losses were switched off. 
Then the interference term must disappear together with the second dip. 
The numerical
calculations have confirmed this prediction for different $\gamma_g$ 
and $E_{\rm max}$.

\begin{figure}[ht]
\includegraphics[width=8.0cm]{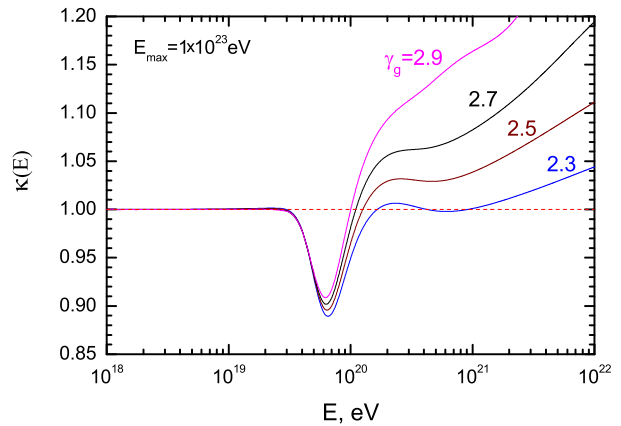}
\caption{The distortion factor $\kappa$ as function of $E$ for 
$E_{\rm max}=1\times 10^{23}$~eV and different values of the spectral
index $\gamma_g$. 
}
\label{gamma}
\end{figure}

In Fig.~\ref{gamma}, the distortion factor is shown for different
values of $\gamma_g$. One may observe the universality of the second dip
with respect to variations of the spectral index, as expected from the
triggering mechanism. We have performed these calculations using the
FP equation. 

Figures~1 and ~2 show that the second dip is not sensitive to the
exact values of $\gamma_g$ and $E_{\rm max}$. This implies also
that a distribution of  $\gamma_g$ and $E_{\max}$ values does not
change the shape and position of the dip. Moreover, the cosmological
evolution of sources is negligible at the energy of the second dip. 
The presence of nuclei primaries affects the second dip only 
for an extreme assumption about the fraction of nuclei. 
Light nuclei are photo-disintegrated at this energy and only the
heaviest nuclei like  Al and Fe survive. Their fraction at the production
should be higher than 20\% in order to hide the second dip~\cite{mod}. 

\begin{figure}[t]
\includegraphics[width=8.0cm]{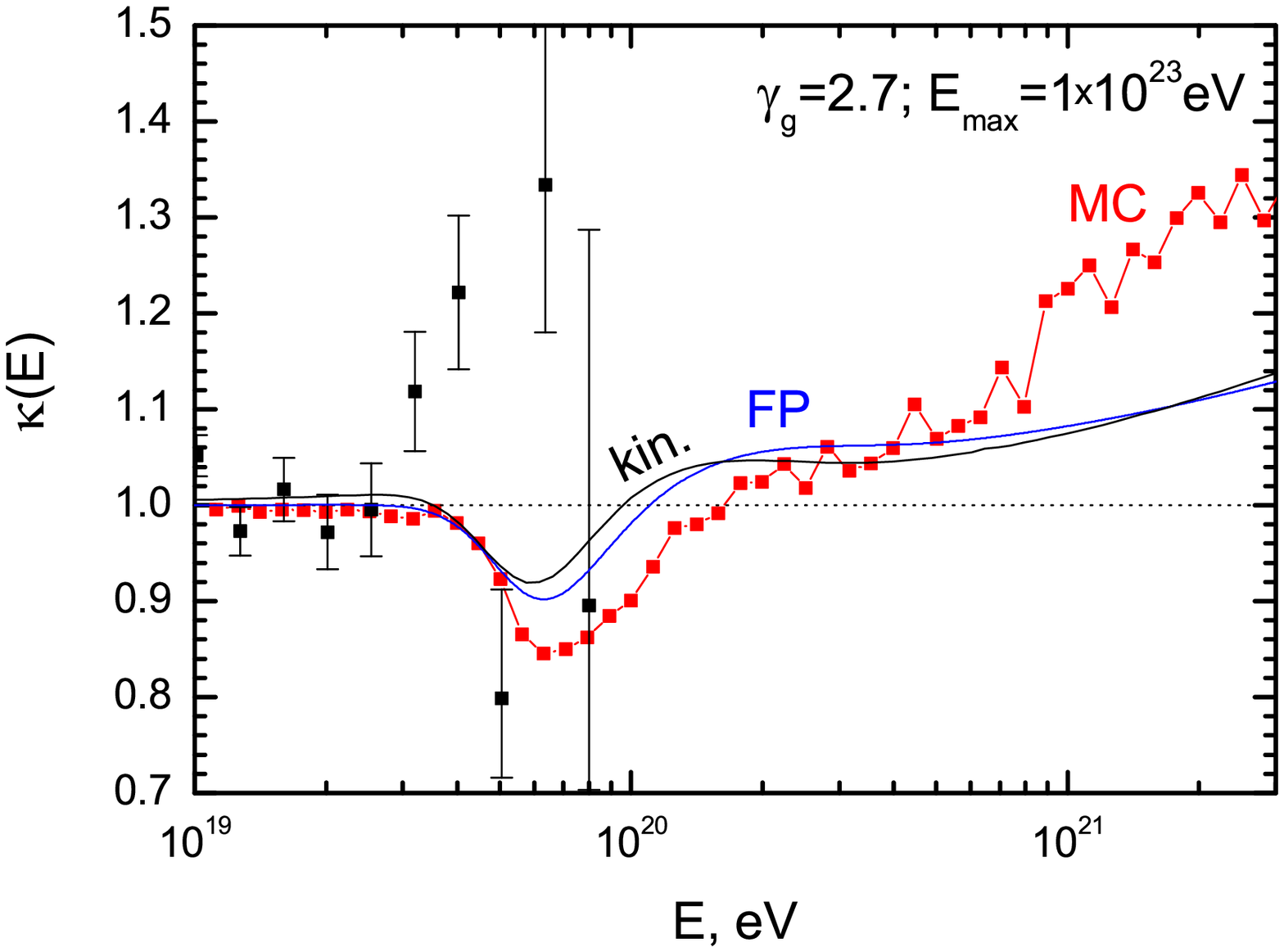}
\includegraphics[width=8.3cm]{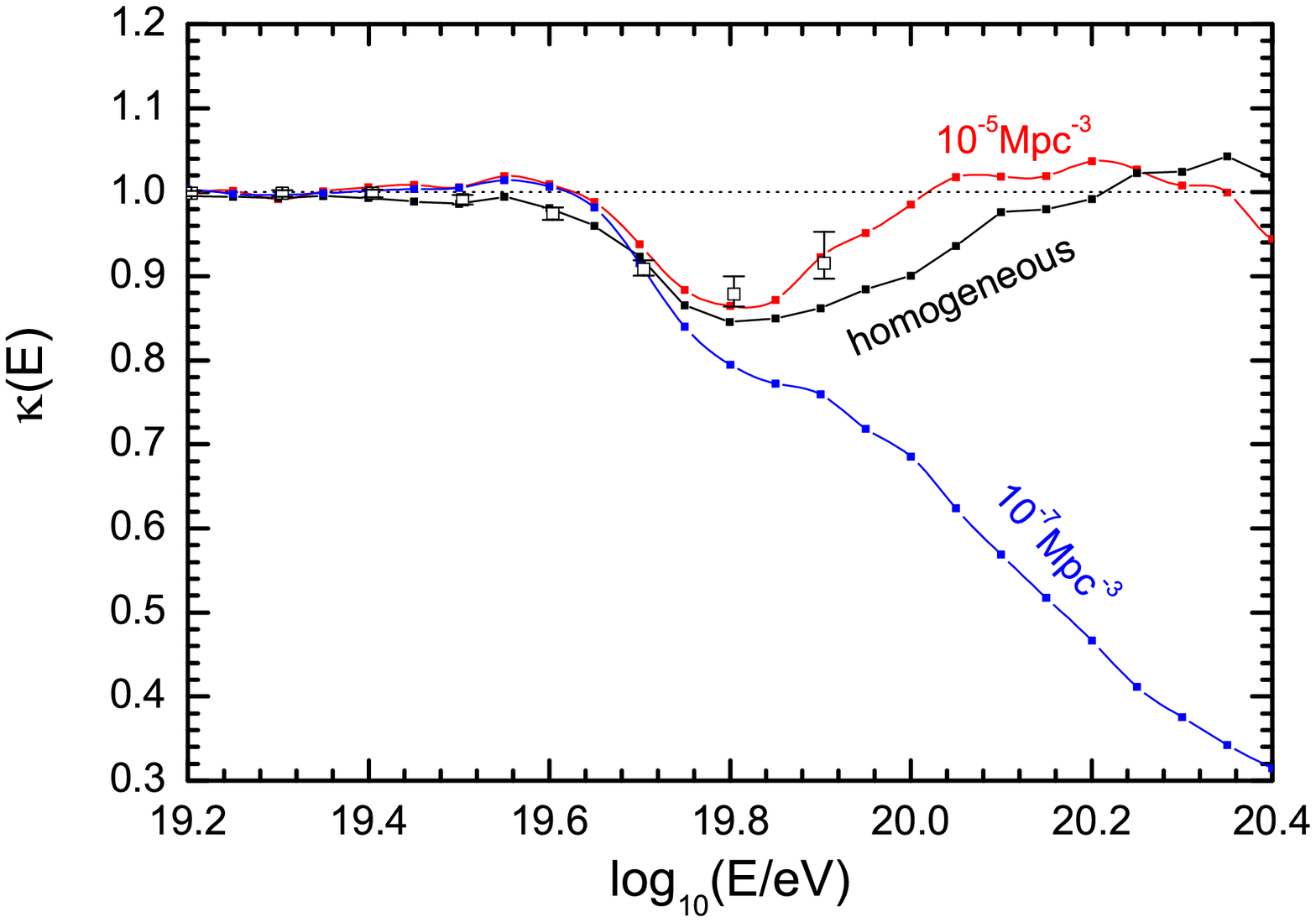}
\caption{Top: The distortion factor calculated for $E_{\rm max}=1\times
  10^{23}$~eV and  $\gamma_g =2.7$ using the kinetic equation (curve
  'kin'), the Fokker-Planck equation (curve 'FP') and Monte Carlo
  calculations 
  (curve 'MC') for a homogeneous source distribution
   together the AGASA data with error bars 
  reduced by a factor 3.  
  Bottom: The distortion factor for a discrete 
  source distribution, 
  $n_s=10^{-5}~{\rm Mpc}^{-3}$, inspired by small-scale
  clustering, and  $n_s=10^{-7}~{\rm Mpc}^{-3}$.}
  \label{comparison}
\end{figure}

As our final test for the second dip we calculate the spectrum with a
Monte Carlo (MC) simulation. The result of a  MC simulation must coincide
with the solution of the kinetic equation, if all relevant parameters
of the problem are identical and when the number of MC runs
tends to infinity. In our case, equal conditions means a
homogeneous source distribution, the same generation
spectra and  $E_{\rm max}$ as well as identical $p\gamma$ interactions. 
We run the Monte Carlo simulation as described in Ref.~\cite{KS} using 
SOPHIA~\cite{Sophia} for the photo-pion interactions, while
in the kinetic equation approach the calculations from Ref.~\cite{BGG}
were used.
As long as only average energy losses are concerned, the results
of both works coincide very well (see \cite{BGG} for a
comparison). However, already small differences in the modeling of
(differential)  cross sections of order of a few percent can result in sizable
variations of the  distortion factor $\kappa$. Numerical errors in the
calculations are another source of possible discrepancies.  
In Fig.~\ref{comparison} (top), we compare the distortion factors
calculated with the kinetic equation, FP equation and MC 
methods for an homogeneous source density. 
The narrow second dip with minimum at 
$E_{\rm 2dip}=6.3\times 10^{19}$~eV is present in all calculations
with small differences in shapes. The points from MC simulation
are connected by straight lines, which helps to see the statistical
uncertainties present especially at high energies.

We have also performed MC calculations for a discrete distribution of
the sources using the values $n_s=10^{-5}~{\rm Mpc}^{-3}$
inspired by small-angle clustering and the very low density 
$n_s=10^{-7}~{\rm Mpc}^{-3}$, both shown in the bottom panel of
Fig.~\ref{comparison}. In the first case, the dip
agrees well with those shown in the upper panel, while
in the latter case the large distance $\sim 200$~Mpc to the nearest
sources results in an early, very steep GZK cutoff that covers up the
second dip. 

We have compared these calculations with the AGASA
data~\cite{AG}.  The experimental distortion
factors are obtained dividing the observed flux by the universal
flux and normalizing the distortion factor at low energies to 
$\kappa = 1$. 
We have diminished the true AGASA error bars by a factor three to
give an impression of the potential of the Pierre Auger Observatory
(PAO) to observe the second dip. This factor corresponds 
to a factor 10 improvement in statistics of PAO compared to AGASA. 
Even if the two data points at $4\times 10^{19}$~eV and $6\times
10^{19}$~eV would lie exactly on the predicted dip (this is
quite possible for the true AGASA error bars), the large  
error bars in the PAO data will prevent a reliable conclusion
on the presence of the second dip. The second dip is expected
to be seen in the future JEM-EUSO space experiment~\cite{EUSO}, 
which will have a 100 times higher  statistics than Auger 
(see bottom panel of  Fig.~\ref{comparison}).
However, this expectation depends critically on the final 
energy threshold of this experiment, which is currently estimated 
as $5\times 10^{19}$~eV but is planned to be lowered~\cite{EUSO}.
The second dip may be used as energy calibrator for this experiment, 
but accurate MC detector simulations are needed for this conclusion.

{\em Conclusions.---}%
We have found a new signature of the interactions of extragalactic UHE
protons with the CMB radiation---the second dip. It is explained by
the interference of pair and photo-pion production and has the shape
of a narrow and shallow dip. 
The second dip is not seen if the admixture of nuclei heavier
than Al in the generation flux is larger than 20\% and if the space
density of the sources is extremely small.
The observation of the second dip is challenging for present 
experiments such as the PAO and requires future high-statistic cosmic ray
detectors. Combined with the steepening of the UHECR  spectrum, its 
observation provides an unambiguous signature of the GZK effect.  

{\em Acknowledgments---}%
We gratefully acknowledge the participation of Svetlana Grigorieva at
the early stage of this work.
We thank ILIAS-TARI for access to the LNGS research infrastructure and
for financial support through EU contract RII-CT-2004-506222.


\end{document}